\documentclass[useAMS,usenatbib]{mn2e}

\usepackage{textcomp}
\usepackage{amsmath}
\usepackage{bm}
\usepackage{graphicx}
\usepackage{gensymb}
\usepackage{subfig}

\newcommand{\Minus}{\text{\texttt{-}}}
\newcommand{\Plus}{\text{\texttt{+}}}

\newcommand{\hoMpc}{h\,\mathrm{Mpc}^{\Minus 1}}
\newcommand{\unitz}{\widehat{\bm{z}}}
\newcommand{\compi}{\mathrm{i}}
\newcommand{\euler}{\mathrm{e}}

\title[Geometric Bias]{Geometric Biases in Power-Spectrum Measurements}

\author[L. Samushia et al.]{L. Samushia$^{1,2,3}$\thanks{E-mail:
lado@phys.ksu.edu}, E. Branchini$^{4,5,6}$, and W. J. Percival$^{3}$\\
$^{1}$Kansas State University, Manhattan KS 66506, USA\\
$^{2}$National Abastumani Astrophysical Observatory, Ilia State University, 2A
Kazbegi Ave., GE-1060 Tbilisi, Georgia\\
$^{3}$Institute of Cosmology \& Gravitation, Dennis Sciama Building, University
of Portsmouth, Portsmouth, PO1 3FX, UK\\
$^{4}$Department of Physics, Universit`a Roma Tre, Via della Vasca Navale 84, Rome 00146, Italy\\
$^{5}$INFN Sezione di Roma 3, Via della Vasca Navale 84, Rome 00146, Italy\\
$^{6}$INAF, Osservatorio Astronomico di Roma, Monte Porzio Catone, Italy}

\begin{document}

\date{\today}

\pagerange{\pageref{firstpage}--\pageref{lastpage}} 

\maketitle

\label{firstpage}

\begin{abstract}
  The observed distribution of galaxies has local transverse isotropy
  around the line-of-sight (LOS) with respect to the observer. The
  difference in the statistical clustering signal along and across the
  line-of-sight encodes important information about the geometry of
  the Universe, its expansion rate and the rate of growth of structure
  within it. Because the LOS varies across a survey, the standard Fast
  Fourier Transform (FFT) based methods of measuring the Anisotropic
  Power-Spectrum (APS) cannot be used for surveys with wide
  observational footprint, other than to measure the monopole
  moment. We derive a simple analytic formula to quantify the bias for
  higher-order Legendre moments and we demonstrate that it is scale
  independent for a simple survey model, and depends only on the
  observed area. We derive a similar numerical correction formula for
  recently proposed alternative estimators of the APS that are based
  on summing over galaxies rather than using an FFT, and can therefore
  incorporate a varying LOS. We demonstrate that their bias depends on
  scale but not on the observed area. For a quadrupole the bias is
  always less than 1 per cent for $k > 0.01\hoMpc$ at $z > 0.32$. For
  a hexadecapole the bias is below 5 per cent for $k > 0.05\hoMpc$ at
  $z > 0.32$.
\end{abstract}


\section{Introduction}

The three-dimensional Power Spectrum (PS) of galaxies is one of the most
important measurements that can be made from galaxy surveys. The Baryon Acoustic
Oscillation feature in the PS can be used to obtain sub percent constraints on
the expansion history of the Universe; The Redshift-Space Distortions
\citep[RSD;][]{1987MNRAS.227....1K} allow precise measurements of the growth rate
of structure; And the Alcock-Paczynski \citep[AP;][]{1979Natur.281..358A} effect
constraints very tightly the geometry of the Universe \citep[for the most recent
measurements see, e.g.][]{2014MNRAS.441...24A,2014MNRAS.439.3504S}.

Both RSD and AP are imprinted into galaxy distribution as a signature
along the line-of-sight (LOS) from the observer. To extract these
signals in an unbiased way it is important that we analyse the data
using the correct LOS that varies from a galaxy-pair to a galaxy-pair. It is geometrically impossible to
define a Cartesian coordinate grid in such a way that the
$\unitz$-axis is everywhere aligned with the LOS direction. Thus, the
APS cannot be measured in Cartesian coordinates with one of the
directions in dual Fourier space identified with the LOS.

Because the PS is a 2-point statistic, it relies on the properties of pairs of
overdensities, although estimation methods can instead be based on pairs of
galaxies. The varying LOS means that, for pairs of galaxies separated by
wide-angles, the RSD for the galaxies in a pair will not be parallel. The
resulting clustering signal including these wide-angle (WA) effects can be accurately
modelled \citep{1998ApJ...498L...1S,2004ApJ...614...51S}, but the difference
beyond assuming a single LOS to the mid-point between the pair of galaxies is
small \citep{2012MNRAS.420.2102S,2012MNRAS.423.3430B,2015MNRAS.447.1789Y}. In
this paper we will concentrate on quantifying the effect of different methods to
allow for the varying LOS between different pairs of galaxies. We will assume
that the WA effects are small.

For distant surveys covering a small spatial region, the LOS will not
vary significantly across the survey. In order to see where the
approximation of a single-LOS breaks down, suppose that we consider
making a general FT of the overdensity field,
\begin{equation}
\label{eq:deltak}
\tilde{\delta}(\bm{k}) = 
\displaystyle\int\!\mathrm{d}^3r\,\delta(\bm{r})\euler^{-\compi\bm{k}\bm{r}},
\end{equation}
\noindent
using an FFT algorithm, with $\unitz$-axis pointed
towards the middle of the survey. The APS could then be computed by averaging
$\tilde{\delta}(\bm{k})$ in wavenumber bins,
\begin{equation}
\label{eq:pk}
P(\bm{k}) = \frac{1}{V}\left<\tilde{\delta}(\bm{k})\tilde{\delta}^*(\bm{k})\right>,
\end{equation}
\noindent
where $V$ is the volume of the survey.  This method has been
successfully used in the past to measure the monopole moment of the
power spectrum, where the LOS direction is irrelevant, and we will now
contrast it with recent methods for measuring higher-order moments. We
will refer to this method as a \textit{single-LOS} method.

What we actually want is a method to compute the APS, or moments of
it, as if there were no LOS variations. i.e. automatically correct for
LOS variations in the computation of the APS, allowing the fast
comparison with models and the retention of all information.  The most
natural way of doing this, allowing for a radially-orientated LOS is
to perform an FT in spherical coordinates with the origin at the
observers position. Even though approach has been applied to previous
galaxy surveys \citep[see,
e.g.][]{1999MNRAS.305..527T,2004MNRAS.353.1201P,2012A&A...540A..60L},
FFTs have an advantage of being faster and more efficient in terms of
computing both the PS estimators and the subsequent likelihood
analysis of the measurements.

In principle, it is also possible to measure the corrected APS on a
Cartesian grid by replacing the FFT with a sum over galaxy-pairs
\citep{2006PASJ...58...93Y}. This allows a LOS to be defined for each
galaxy pair in the sample (effectively having a separate coordinate
grid, or $\unitz$-axis for each galaxy pair). It is however
impractical for modern large galaxy surveys, as it would require
prohibitively large computing times. We will refer to this method as a
\textit{pairwise-LOS} method (see Sec.~\ref{ssec:pgrid}). As discussed
above, in this paper we will ignore WA effects, and thus
consider that the pairwise-LOS method gives an exact result.

\begin{figure*}
\subfloat[]{\includegraphics[width=0.5\textwidth]{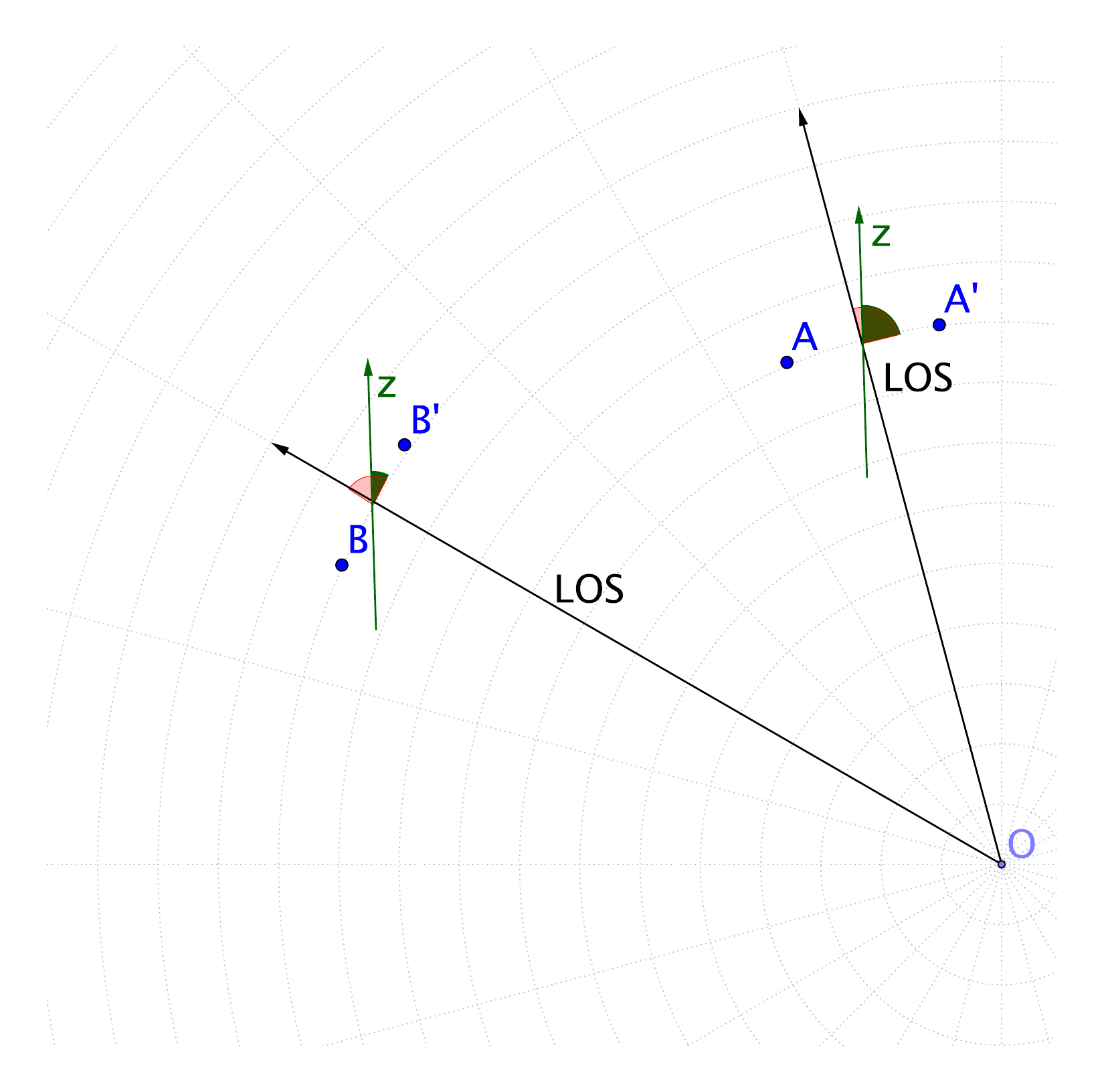}}
\subfloat[]{\includegraphics[width=0.5\textwidth]{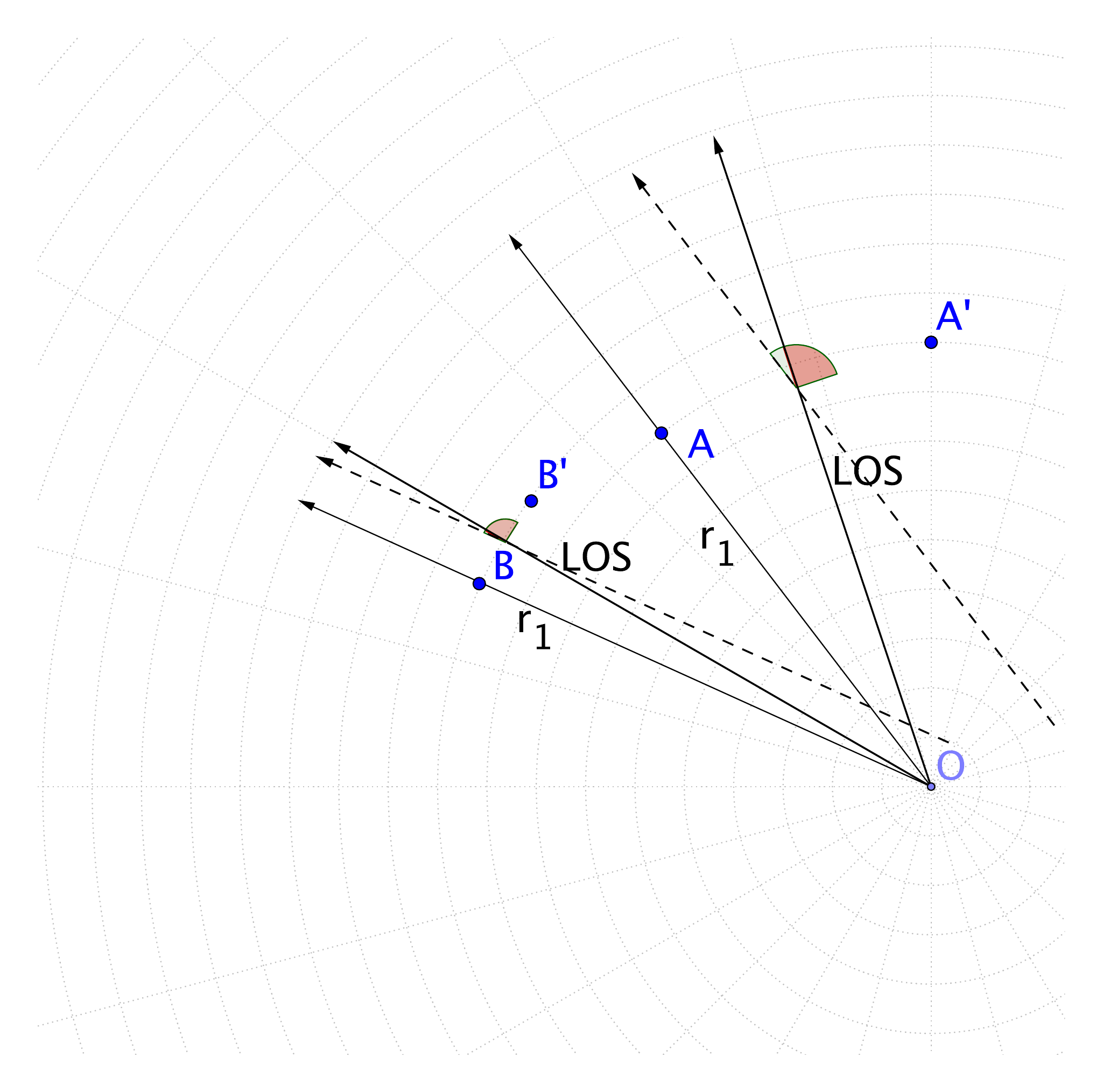}}
\caption{(a) Angles between the pair separation vector and the LOS assumed in
the pairwise-LOS method (red) and the single-LOS method (green). (b) Angles
between the pair separation vector and the LOS assumed in the pairwise-LOS
method (red) and the moving-LOS (green) methods, for two example pairs of
galaxies AA' and BB'. The moving-LOS method fails (i.e. red and green angles are
significantly different) for pairs with large separations, while the single-LOS
method fails for pairs whose true LOS is different from the fixed LOS assumed in
the single-LOS method. The dashed arrows on panel (b) are parallel to lines $OB$
and $OA$.}
\label{fig:approx}
\end{figure*}

A practical approximation for the pairwise-LOS method is to define a LOS for a
chosen galaxy in each pair. This allows the estimation method to be reduced from
a sum over pairs to a sum over galaxies, which is computationally faster
\citep{2011MNRAS.415.2876B} . This approximation will break down for galaxy
pairs with very large angular separation (it is effectively another WA effect)
but will become increasingly accurate for smaller scale measurements. The
algorithm can be reduced to series of FFTs
\citep{2015arXiv150505341B,2015arXiv150602729S} and is significantly faster than
the pairwise-LOS algorithms, which makes it feasible for the analysis of galaxy
surveys.  This method has been applied to WiggleZ data
\citep{2011MNRAS.415.2876B} and BOSS data \citep{2014MNRAS.443.1065B} and the
results suggest that there isn't an appreciable bias with respect to the
pairwise-LOS method. We will refer to this method as a \textit{moving-LOS}
method (see Sec.~\ref{ssec:mgrid}).

In this paper we aim to quantify the biases in the APS induced by
single-LOS and moving-LOS methods with respect to the pairwise-LOS
method. To make the discussion clearer we make certain simplifying
assumptions: We ignore the effects of mask and selection function
(window effects). Correcting for the mask and selection effects is not
a trivial task, but this problem is almost independent from the issue
that we want to address in this work, so we will assume that the
window effects have been properly dealt with to required accuracy. We
also ignore the discrete nature of galaxy survey data and will write
all equations as integrals over overdensity field rather than sums
over galaxies. These assumptions help to keep the discussion clearer
and the equations compact and don't affect any of our main
conclusions. The approximation is further justified by the fact that
for large separations (where the geometric bias is larger) the effects
of discreteness of the galaxy field are negligible.

We denote vectors by bold symbols ($\bm{r}$), unit vectors by bold
symbols with a hat ($\widehat{\bm{r}}=\bm{r}/|r|$), and the modulus of
a vector with italic symbols ($r = \bm{r}\bm{r}/|r|^2$). A scalar
product is assumed between two sequential (unit) vectors.

\section{Anisotropic Power-Spectrum}
\label{sec:aps}

We start with the basic premise of a correlated galaxy overdensity
field, with the correlation function (CF) between two galaxies at
positions $\bm{r}_1$ and $\bm{r}_2$,
$\xi(\bm{r}_1,\bm{r}_2)\equiv\left<\delta(\bm{r}_1)\delta(\bm{r}_2)\right>$. For
convenience, we define two vectors
\begin{eqnarray}
  \bm{r}_\Minus &\equiv& \bm{r}_2 - \bm{r}_1,\\
  \bm{r}_\Plus &\equiv& (\bm{r}_2 + \bm{r}_1)/2,
\end{eqnarray}
where $\bm{r}_\Minus$ connects the two galaxies and $\bm{r}_\Plus$ is
the vector from the observer to to their midpoint, which we will
identify with the LOS of the galaxy pair. Because of local transverse
isotropy around the line-of-sight, and our assumption of no WA
effects, the CF will only be a function of the distance between the
galaxies and the angle they make with respect to the
LOS. $\xi(\bm{r}_1,\bm{r}_2) =
\xi(r_\Minus,\widehat{\bm{r}}_\Minus\widehat{\bm{r}}_\Plus)$.

The angular dependence of the CF is usually expanded into Legendre polynomials
\begin{equation}
\xi(r_\Minus,\widehat{\bm{r}}_\Minus\widehat{\bm{r}}_\Plus) =
\displaystyle\sum_\ell\xi_\ell(r_\Minus)\mathcal{L}_\ell(\widehat{\bm{r}}_\Minus\widehat{\bm{r}}_\Plus),
\end{equation}
\noindent
with most of the useful information in first three even multipoles
\citep{2011PhRvD..83j3527T,2012MNRAS.419.3223K}.

The APS is defined as a FT of CF and can also be decomposed into Legendre
polynomials with respect to LOS
\begin{equation}
\label{eq:pkl1}
P(\bm{k}) \equiv 
\displaystyle\int\!\mathrm{d}^3r_\Minus\,\xi(r_\Minus,\widehat{\bm{r}}_\Minus\widehat{\bm{r}}_\Plus)\euler^{-\compi\bm{k}\bm{r}_\Minus}
=
\displaystyle\sum_\ell P_\ell(k)\mathcal{L}_\ell(\widehat{\bm{k}}\widehat{\bm{r}}_\Plus).
\end{equation}
\noindent
This is a standard definition and theoretical predictions of APS are computed
for this quantity \citep[see e.g.,][and references
therein]{2015arXiv150606596R}.
The PS multipoles are related to the CF multipoles by
\begin{equation}
\label{eq:pkl2}
P_\ell(k) = 4\pi \compi^\ell 
\displaystyle\int\!\mathrm{d}r_\Minus\,\xi_\ell(r_\Minus)j_\ell(kr_\Minus)r_\Minus^2.
\end{equation}

\subsection{Pairwise-LOS Method}
\label{ssec:pgrid}

In the pairwise-LOS method one would correct for the varying LOS, by
computing the integral over the overdensity field simultaneously
assigning correct LOS direction to all galaxy pairs. In other words,
one would compute the multi-dimensional integral
\begin{align}
\nonumber
P_\ell(k) = \frac{2\ell+1}{4\pi V} &\displaystyle\int\!\mathrm{d}\widehat{\bm{k}}\mathrm{d}^3r_1\mathrm{d}^3r_2 \\
&\delta(\bm{r}_1)\delta(\bm{r}_2)\euler^{-\compi\bm{k}\bm{r}_1}
\euler^{\compi\bm{k}\bm{r}_2}\mathcal{L}_\ell(\widehat{\bm{k}}\widehat{\bm{r}_\Plus}).
\label{eq:Plmg}
\end{align}
\noindent
The expectation value of this integral is
\begin{align}
\label{eq:pgrid}
\nonumber
P_\ell(k) = \frac{2\ell+1}{4\pi V} \displaystyle\int\!\mathrm{d}\widehat{\bm{k}}&\mathrm{d}^3r_\Minus\mathrm{d}^3r_\Plus\\
&\xi(r_\Minus,\widehat{\bm{r}}_\Minus\widehat{\bm{r}}_\Plus)\euler^{-\compi\bm{k}\bm{r}_\Minus}\mathcal{L}_\ell(\widehat{\bm{k}}\widehat{\bm{r}_\Plus})
\end{align}
\noindent
and the APS computed in such way would coincide with the definitions of
Eq.~(\ref{eq:pkl1}) and~(\ref{eq:pkl2}) and would therefore be exact. Computing
this multi-dimensional integral over large galaxy sample however demands prohibitively
large CPU time and is not currently viable. 

\subsection{Single-LOS Method}
\label{ssec:fgrid}
In the single-LOS method the $\unitz$-axis is pointed towards the middle of the
survey footprint and it's assumed that $\widehat{\bm{r}}_\Plus \sim \unitz$
within the survey volume. This approximation allows us to rewrite
Eq.~(\ref{eq:Plmg}) as
\begin{equation}
P_\ell(k) = \frac{2\ell+1}{4\pi V}
\displaystyle\int\!\mathrm{d}\widehat{\bm{k}}\mathcal{L}_\ell(\widehat{\bm{k}}\unitz)
\left|\displaystyle\int\!\mathrm{d}^3\bm{r}\delta(\bm{r})\euler^{-\compi\bm{k}\bm{r}}\right|^2.
\end{equation}
The expectation value of the integral is
\begin{equation}
P_\ell(k) = \frac{2\ell + 1}{2}\frac{1}{4\pi}
\displaystyle\int\!\mathrm{d}\widehat{\bm{k}}\,P(\bm{k})\mathcal{L}_\ell(\widehat{\bm{k}}\unitz),
\end{equation}
\noindent
where
\begin{equation}
P(\bm{k}) = \frac{1}{V}
\displaystyle\iint\!\mathrm{d}^3r_\Minus\mathrm{d}^3r_\Plus\xi(r_\Minus,
\widehat{\bm{r}}_\Minus\widehat{\bm{r}}_\Plus) \euler^{-\compi\bm{k}\bm{r}_\Minus},
\end{equation}
as before. After integrating over $\bm{r}_\Minus$ and $\widehat{\bm{k}}$ this
expression reduces to
\begin{equation}
\label{eq:fgridbias}
P_\ell(k) =
P_\ell^\mathrm{t}(k)\frac{1}{V}\displaystyle\int\!\mathrm{d}^3r_\Plus\mathcal{L}_\ell(\unitz\widehat{\bm{r}}_\Plus),
\end{equation}
\noindent
where $P^\mathrm{t}$ is the true PS multipole defined by Eqs.~(\ref{eq:pkl1}) and~(\ref{eq:pkl2}).

For $\ell = 0$ the single-LOS APS reduces to the true APS as $\mathcal{L}_0(x) =
1$ -- The monopole of the APS is unbiased even when measured with the single-LOS
method. For higher order multipoles the bias is always present, but could be
small in some limits. For example, if the survey area is small and the
$\unitz$-axis is pointed towards the center of the survey, LOS directions of all
pairs will be very close to $\unitz$. In this case, $\widehat{\bm{r}}_\Plus \sim
\unitz$ and $\mathcal{L}_\ell(\unitz\widehat{\bm{r}}_\Plus) \sim 1$, resulting
in small bias.

\subsection{Moving-LOS Method}
\label{ssec:mgrid}

The moving-LOS method is an approximation of Eq.~(\ref{eq:pgrid}). The
overdensity field is transformed as
\begin{equation}
\label{eq:mgrid1}
\tilde{\delta}_\ell(\bm{k}) = 
\displaystyle\int\!\mathrm{d}^3r\,\delta(\bm{r})\mathcal{L}_\ell(\widehat{\bm{k}}\widehat{\bm{r}})\euler^{-\compi\bm{k}\bm{r}}
\end{equation}
\noindent
and the APS multipoles are computed as
\begin{equation}
\label{eq:mgrid2}
P_\ell(k) = \frac{2\ell + 1}{4\pi V}
\displaystyle\int\!\mathrm{d}\widehat{\bm{k}}\tilde{\delta}_0(\bm{k})\tilde{\delta}_\ell(\bm{k}).
\end{equation}

The APS computed in using this expression is equivalent to
\begin{align}
\nonumber
P_\ell(k) = \frac{2\ell + 1}{4\pi V} &\displaystyle\int\!\mathrm{d}^3r_1\mathrm{d}^3r_2\mathrm{d}\widehat{\bm{k}}\\
&\delta(\bm{r}_1)\delta(\bm{r}_2)\euler^{-\compi\bm{k}\bm{r}_1}\euler^{\compi\bm{k}\bm{r}_2}\mathcal{L}_\ell(\widehat{\bm{k}}\widehat{\bm{r}_1})
\end{align}
\noindent
and the expectation value of that expression is
\begin{align}
\label{eq:mgridexp}
\nonumber
P_\ell(k) = \frac{2\ell + 1}{4\pi V} \displaystyle\int\!\mathrm{d}^3{r}_\Minus&\mathrm{d}^3r_\Plus\mathrm{d}\widehat{\bm{k}}\\
&\xi(r_\Minus,\widehat{\bm{r}}_\Minus\widehat{\bm{r}}_\Plus)\euler^{-\compi\bm{k}\bm{r}_\Minus}\mathcal{L}_\ell(\widehat{\bm{k}}\widehat{\bm{r}_1}).
\end{align}
\noindent
Using properties of spherical harmonics this can be further reduced to
\begin{equation}
\label{eq:mpl}
P_\ell(k) =
\displaystyle\int\!\mathrm{d}k'\displaystyle\sum_{\ell'}P^\mathrm{t}_{\ell'}(k')W_{\ell\ell'}(k,
k'),
\end{equation}
\noindent
where
\begin{equation}
W_{\ell,\ell'}(k,k') = \frac{2k'^2}{\pi}\displaystyle\int\!\mathrm{d}r_\Minus
r_\Minus^2
j_\ell(kr_\Minus)j_{\ell'}(k'r_\Minus)\mathcal{F}_{\ell\ell'}(r_\Minus)
\end{equation}
\noindent
and
\begin{align}
\label{eq:Fll}
\mathcal{F}_{\ell,\ell'}(r_\Minus) \equiv &\frac{4\pi\mathrm{i}^{\ell-\ell'}}{V(2\ell'+1)}\displaystyle\int\!\mathrm{d}^3r_\Plus\mathrm{d}\widehat{\bm{r}}_\Minus\\
\nonumber
\displaystyle\sum_{m m'} &Y_{\ell' m'}(\widehat{\bm{r}}_\Minus)Y^*_{\ell' m'}(\widehat{\bm{r}}_\Plus)Y^*_{\ell m}(\widehat{\bm{r}}_\Minus)Y_{\ell m}(\widehat{\bm{r}}_1)
\end{align}
\noindent
(See App.~\ref{app:Fll} for details).

This expression will reduce to the true APS only if
$W_{\ell\ell'}(k,k')=\delta_{\ell\ell'}\delta(k - k')$.  In
the limit of very small separations $\widehat{\bm{r}}_\Plus \sim
\widehat{\bm{r}}_1$, the $\mathcal{F}$ tends to $\delta_{\ell\ell'}$ by the
virtue of the closure relation for the spherical harmonics, making
$W_{\ell\ell'}(k,k')$ converge to $\delta_{\ell\ell'}\delta(k
- k')$ in the mean.\footnote{In this limit the moving-LOS method basically reduces to the
pairwise-LOS, since the pairwise-LOS method will have the same expansion as
Eq.~(\ref{eq:Fll}) but with $\bm{r}_1$ replaced by $\bm{r}_\Plus$.} Properties of spherical harmonics also enforce the condition
  $W_{0\ell'} \propto \delta_{0\ell'}$ for all $k$ and $k'$, which means that,
as with the single-LOS method, there is no bias for $\ell=0$.  Unlike the
single-LOS method, the bias can't be expressed as a simple scale-independent
scaling of the APS.

\section{APS bias as a function of sky area, redshift, and scale}

In this section we will quantify the biases in the single-LOS and
moving-LOS methods, using the pairwise-LOS method as the reference.
To compute the biases we need to specify the geometry of the observed
volume as the biases will depend on the distribution of pair
separations. For simplicity, we will assume that the observed patch of
the sky is circularly symmetric, the $\unitz$-axis is pointing to the
center of the observed area (this choice results in the least bias for
the single-LOS method with the assumed LOS along this direction), and
the mask and selection functions are uniform. We will also assume that
the width of redshift bin is small compared to the distance of the
sample from the observer. This simple model clearly lacks the detailed
window of an actual survey, but the angular distribution of pair
separations will be roughly correct, and the radial thinness is a
conservative choice as it forces pair separations of the same physical
separation to wider angular separations.

\begin{figure}
\includegraphics[width=0.5\textwidth]{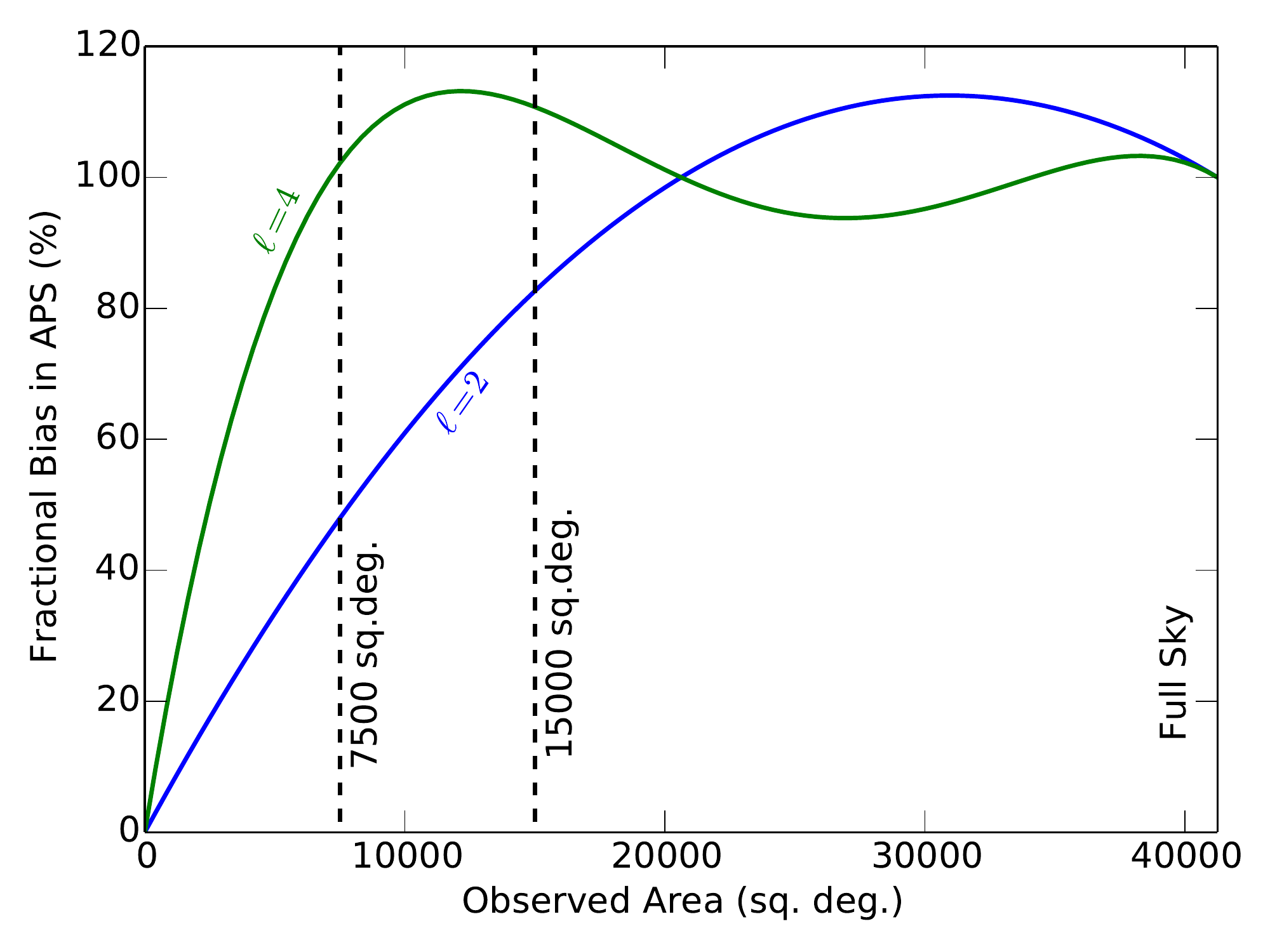}
\caption{Fractional error in the measured APS as a function of observed area for the
single-LOS method.}
\label{fig:fgridbias}
\end{figure}

For this geometry, using a single-LOS analysis the bias, given by Eq.~(\ref{eq:fgridbias}), reduces to
\begin{equation}
P_\ell(k) =
P_\ell^\mathrm{t}(k)\frac{\displaystyle\int_0^{\vartheta_\mathrm{max}}\!\!\!\!\!\!\!\!\!\!\!\mathrm{d}\vartheta\,\sin(\vartheta)\mathcal{L}_\ell[\cos(\vartheta)]}{\displaystyle\int_0^{\vartheta_\mathrm{max}}\!\!\!\!\!\!\!\!\!\!\!\mathrm{d}\vartheta\,\sin(\vartheta)}.
\end{equation}
\noindent
The bias in the APS is independent of the wavenumber and redshift and
only depends on the angular extent of the observed area. We plot this
bias for $\ell = 2$ and $\ell=4$ in Fig.~\ref{fig:fgridbias}. We see a
gradual increase in the biases with area for small surveys, with
biases that are already larger than 1 per cent when the footprint is
only 1000 $\mathrm{deg}^2$. For hemispherical or full-sky surveys with
a single-LOS, both the quadrupole ($\ell=2$) and the hexadecapole
($\ell=4$) are zero (the fractional error is 1): here any increase in
the clustering strength caused by RSD along the single-LOS is matched
by an increase perpendicular to the single-LOS around the edges of each
hemisphere.

For the moving-LOS method, the APS bias given by Eq.~(\ref{eq:mpl})
cannot be expressed as a simple ratio of true and measured
power-spectra. For the simple -- ``thin spherical cap'' -- geometry we
have assumed, we can use the properties of spherical harmonics to
reduce the five dimensional integral in Eq.~(\ref{eq:Fll}) to a one
dimensional integral
\begin{align}
\label{eq:Fllapp}
\mathcal{F}_{\ell\ell'}(\eta) = &\mathrm{i}^{\ell-\ell'}\frac{2\ell + 1}{2} \displaystyle\int_{-1}^1\!\mathrm{d}\mu\,\\ 
&\mathcal{P}_{\ell'}^0(\mu)
\mathcal{P}_{\ell}^0(\mu)\mathcal{P}_{\ell}^0\left(\frac{1-\eta\mu/2}{\sqrt{1-\eta\mu+\eta^2/4}}\right)\nonumber
\end{align}
\noindent
(see, App.~\ref{app:Fll}) where $\mathcal{P}$ denote associated
Legendre polynomials and $\eta \equiv r_\Minus/r_\Plus$. Examining
Eq.~(\ref{eq:Fllapp}) we see that the expression tends to
$\delta_{\ell\ell'}$ when $\eta$ tends to zero, suggesting that, for a
fixed scale, the approximation works better the further away that
galaxies are from the observer, as expected. Eq.~(\ref{eq:Fllapp})
also shows that the bias depends only on the wavenumber and doesn't
depend on the observed sky coverage: given that the bias is only
related to how each pair of galaxies is treated rather than the
distribution of pairs, this is also expected for scales unaffected by
the window.

\begin{figure}
\includegraphics[width=0.5\textwidth]{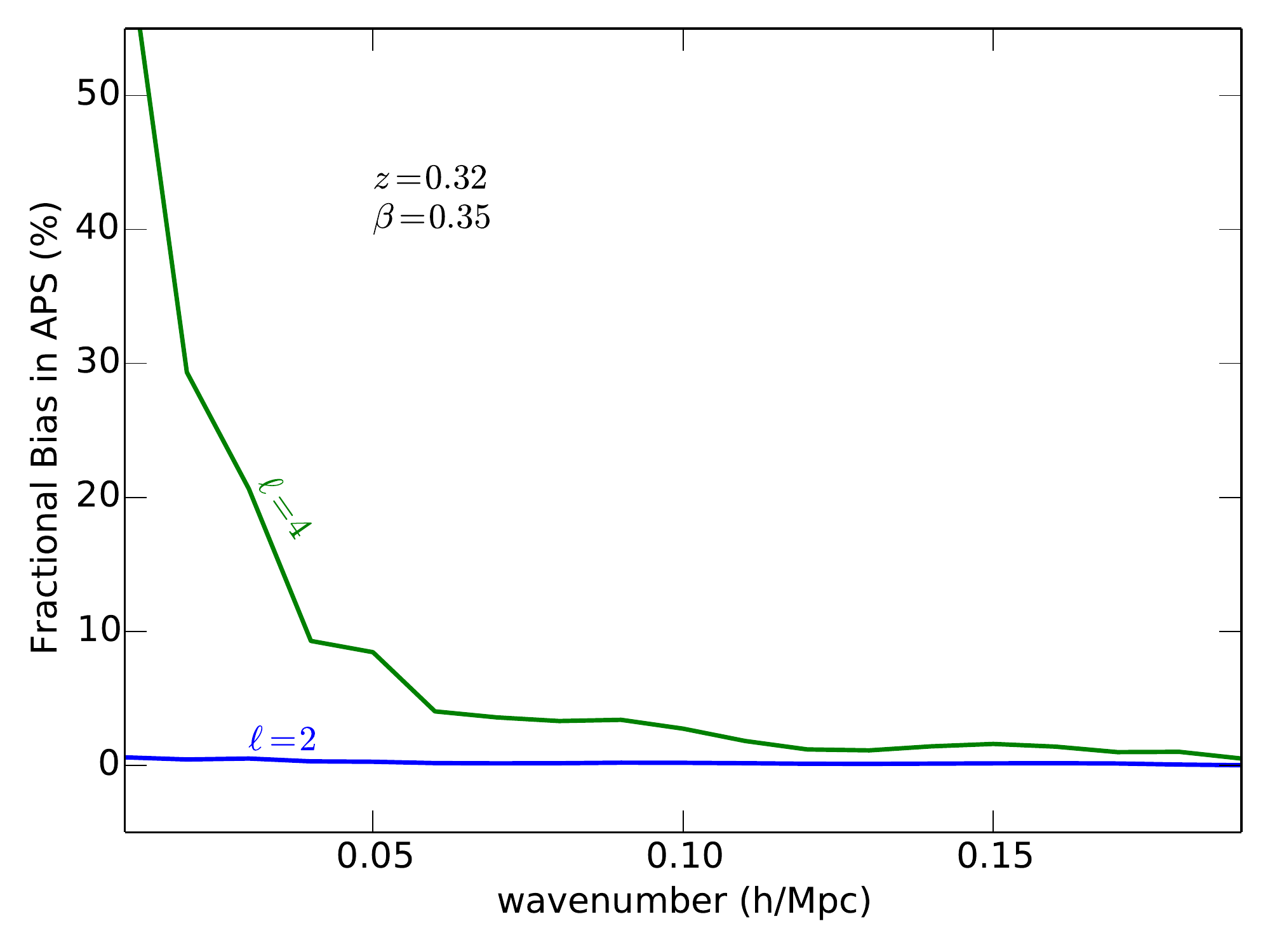}
\caption{Fractional error in the measured APS as a function of observed area for
the moving-LOS method for $z=0.32$ and $\beta=0.35$.}
\label{fig:mgridbias}
\end{figure}

Fig.~\ref{fig:mgridbias} shows the fractional bias in the APS at $z =
0.32$ for $\beta = 0.35$.\footnote{$z=0.32$ is an effective redshift
  of BOSS CMASS sample. $\beta$ parameter describes the amount of
  anisotropy in APS \citep[see, e.g.][for a proper
  definition]{1998ASSL..231..185H}.} The bias in the quadrupole lies
below 1 per cent at wavenumbers above 0.01 $h$/Mpc. The bias in the
hexadecapole is larger and reaches a sub per cent level only for
wavenumbers above 0.1 $h$/Mpc. At higher redshifts the biases are
reduced even further.

\section{Summary and Comparison with Previous Work}
\label{sec:summary}

In this work we quantify biases on the APS measured by various computational
methods, focussing on methods correcting for the varying LOS. We have ignored
wide-angle effects that arise because the peculiar velocity shifts in galaxies
in a pair are not parallel. At 200$h^{\Minus 1}$ Mpc this effect is less than a
per cent already at $z = 0.2$, and decreases for the smaller scales usually of
interest in analyses \citep[see
e.g.][]{2012MNRAS.420.2102S,2010MNRAS.409.1525R}. Instead, the approximations
discussed in this work are related to the way of measuring the APS, and we have
compared two methods - one known to be wrong (single-LOS), and one where small
biases have previously been assumed
\citep{2011MNRAS.415.2876B,2014MNRAS.443.1065B} but that we now quantify
(moving-LOS).

To provide a baseline, we have considered ignoring the variation in the LOS and
assuming a single-LOS: here we have shown that we are left with a significant
bias in measurements other than the monopole moment of the power spectrum, that
is independent of scale and only depends on the area of the angular footprint.
This has been know for a long time, and consequently this method has not been
widely used to measure the APS, other than the monopole. We have shown through
an analytic formula for the induced bias of APS multipoles
(Eq.~\ref{eq:fgridbias}), that the bias on the quadrupole is large even for
surveys as small as 1000\,deg$^2$. Our results can be used to correct this
effect to leading order for analysis methods in which the observed area is
subdivided into smaller regions to make the bias smaller \citep[see
e.g.,][]{1994MNRAS.267..785C,2014MNRAS.445.3737H}.

We have also considered a measurement method proposed more recently,
that performs the transform summing over galaxies in a way that allows
varying LOS to be factored into the measurement. Ideally we would want
to sum over pairs rather than galaxies, but this is impractical, and
this revision leads to a small bias in the anisotropic
measurements. We have derived an analytic formula for this bais that,
in the limit of narrow redshift bin, can be reduced to a one
dimensional integral over the true power-spectrum
(Eq.~\ref{eq:mpl}). This shows that the bias is a function of scale
(but not the area) and only depends on the ratio of pair-separation to
the distance of pair from the observer $r_\Minus/r_\Plus$. While
small, this correction can easily be calculated and measurements
corrected for this effect.

The new method calculating the APS using a sum over galaxies
(moving-LOS) remains accurate for the quadrupole APS at scales above
$k = 0.01h^{-1}$ Mpc even at small redshifts. The bias in the
hexadecapole moment is larger, but also decays at high redshifts. Our
correction formulas in Eq.~(\ref{eq:fgridbias}) and~(\ref{eq:mpl}) can
be used to correct the biases to leading order. 

Our analysis of the significance of making various LOS approximations makes a
number of simplifying assumptions about the survey for the sake of analytical
clarity. We assume the thin-shell approximation and ignore boundary effects, the
effects of mask, and the redshift dependence of the mean galaxy number density.
While the primary effects of these assumptions can be corrected when making
clustering measurements, it is likely that they couple with the geometric biases
considered here. Indeed, these approximations are likely to change the
corrections due to LOS-assumptions at a comparable order to the corrections
presented here for a simplified survey (although when the single-LOS method is
used on a data with wide footprint the geometric effects are likely to be
dominant). A more precise correction would require a detailed study of how the
mask effects couple with the geometric biases.

\section*{Acknowledgements}

LS would like to thank Larry Weaver for useful discussions about the rotation
group and the symmetries of spherical harmonics and Glenn Horton-Smith for
discussions related to the numerical integration of oscillatory integrals. LS is
grateful for support from SNSF grant SCOPES IZ73Z0\_152581. EB is supported by
INFN-PD51 INDARK, MIUR PRIN 2011 and ASI/INAF/I/023/12/0. WJP acknowledges
support from the UK STFC through the consolidated grant ST/K0090X/1, and from
the European Research Council through the Darksurvey grant.

\appendix
\section{Bias in Moving-LOS Method}
\label{app:Fll}
We will rewrite Eq.~(\ref{eq:mgridexp}) using the plane wave expansion,
\begin{equation}
\euler^{\compi\bm{k}\bm{r}} = \displaystyle\sum_{\ell m}i^\ell 4\pi j_\ell(kr)Y_{\ell m}(\widehat{\bm{k}})Y^*_{\ell m}(\widehat{\bm{r}}),
\end{equation}
\noindent
and the addition theorem for Legendre polynomials,
\begin{equation}
\mathcal{L}(\widehat{\bm{k}}\widehat{\bm{r}})=\frac{4\pi}{2\ell+1}\displaystyle\sum_m Y_{\ell m}(\widehat{\bm{k}})Y^*_{\ell m}(\widehat{\bm{r}}).
\end{equation}
This results in
\begin{align} 
\nonumber
P_\ell(k) =& \frac{2\ell+1}{4\pi
V}\displaystyle\sum_{\ell'm'}\displaystyle\sum_{\ell''m''}\displaystyle\sum_{m}\displaystyle\int\!\mathrm{d}^3r_\Minus\mathrm{d}^3r_\Plus\mathrm{d}\widehat{\bm{k}}
\\
\nonumber
&\xi_{\ell'}(r_\Minus)\frac{4\pi}{2\ell'+1}Y_{\ell'm'}(\widehat{\bm{r}}_\Minus)
Y^*_{\ell'm'}(\widehat{\bm{r}}_\Plus)\\
&\compi^{\ell''}4\pi
j_{\ell''}(kr)Y^*_{\ell''m''}(\widehat{\bm{k}})Y_{\ell''m''}(\widehat{\bm{r}}_\Minus)\\
\nonumber
&\frac{4\pi}{2\ell+1}Y_{\ell m}(\widehat{\bm{k}})Y^*_{\ell
m}(\widehat{\bm{r}}_1).
\end{align}
Integrating over $\widehat{\bm{k}}$, by virtue of orthogonality of spherical
harmonics, results in Eqs.~(\ref{eq:mpl})--(\ref{eq:Fll}).

The expression in Eq.~(\ref{eq:Fll}) is invariant under the rotation of
coordinate system. This symmetry can be used to align the direction of
$\widehat{\bm{z}}$ axis with $\widehat{\bm{r}}_\Plus$ as we integrate over
$\mathrm{d}\widehat{\bm{r}}_\Plus$. Ignoring the boundary effects and assuming
that the redshift shell is thin, this results in
in
\begin{align}
\mathcal{F}_{\ell\ell'}(r_\Minus) &=
\mathrm{i}^{\ell-\ell'}\sqrt{\frac{4\pi}{(2\ell'+1)}}\\
\nonumber
&\displaystyle\int\!\mathrm{d}\widehat{\bm{r}}_\Minus
Y_{\ell'0}(\widehat{\bm{r}}_\Minus)\displaystyle\sum_{m}Y^*_{\ell m}(\widehat{\bm{r}}_\Minus)Y_{\ell
m}(\widehat{\bm{r}}_1),
\label{eq:Fllappend}
\end{align}
\noindent
as the integral over $\mathrm{d}^3r_\Plus$ is equal to volume and only terms
with $m'=0$ because $\bm{r}_\Plus$ is pointing in $\widehat{\bm{z}}$ direction.
This expression is again invariant with respect to rotations in azimuthal angle
(ignoring boundary effects) and can be reduced to
\begin{align}
\mathcal{F}_{\ell\ell'}(r_\Minus) &=
2\pi\mathrm{i}^{\ell-\ell'}\sqrt{\frac{4\pi}{(2\ell'+1)}}\\
\nonumber
&\displaystyle\int\!\mathrm{d}\mu_\Minus
Y_{\ell'0}(\mu_\Minus)Y^*_{\ell 0}(\mu_\Minus)Y_{\ell
0}(\mu_1),
\end{align}
\noindent
\noindent
where the $m' \neq 0$ terms are killed by the azimuthal integral and
$\mu_\Minus$ and $\mu_1$ are the cosines of respective polar angles. Since
$\bm{r}_1 = \bm{r}_\Plus - \bm{r}_\Minus/2$ we have $\mu_1 = (1 -
\eta\mu_\Minus/2)/\sqrt{1-\eta\mu+\eta^2/4}$, which will result in
Eq.~(\ref{eq:Fllapp}).

\label{lastpage}

\end{document}